# Giant resonances of endohedral atoms

M. Ya. Amusia[1,2], A. S. Baltenkov[3] and L. V. Chernysheva[2]

[1]Racah Institute of Physics, the Hebrew University, Jerusalem 91904, Israel
[2]Ioffe Physical-Technical Institute, St.-Petersburg 194021, Russia
[3]Arifov Institute of Electronics, Tashkent, 700125, Uzbekistan

**Abstract**

We demonstrate for the first time that the effect of fullerene shell upon photoionization of the "caged" atom in an endohedral can result in formation of Giant Endohedral Resonances or GER. This is illustrated by the concrete case of Xe@$C_{60}$ photoionization cross-section that exhibits at 17 eV a powerful resonance with total oscillator strengths of about 25.

The prominent modification of the $5p^6$ electron photoionization cross-section of Xe@$C_{60}$ takes place due to strong fullerene shell polarization under the action of the incoming electromagnetic wave and oscillation of this cross-section due to the reflection of the photoelectron from Xe by the $C_{60}$. These two factors transform the smoothly decreasing $5p^6$ cross-section of Xe into a rather complex curve with a powerful maximum for Xe@$C_{60}$, with the oscillator strength of it being equal to 25!

We present also the results for the dipole angular anisotropy parameter that is strongly affected by the reflection of the photoelectron waves but not modified by $C_{60}$ polarization.

**1.** In this Letter we consider the photoionization of outer $5p^6$ subshell of Xe endohedral atom, formed by a fullerene $C_{60}$ inside of which a noble gas atom is staffed, Xe@$C_{60}$. Recently, a great deal of attention was and still is concentrated on photoionization of endohedral atoms. It was demonstrated in a number of papers [1-5] that the $C_{60}$ shell action results in prominent resonance structure in the photoionization cross section of endohedral atoms. Although the experimental investigation of A@$C_{60}$ photoionization seems to be very difficult at this moment, it will be inevitably intensively studied in the future[1]. This justifies the current efforts of the theorists that are predicting rather non-trivial effects waiting for verification.

The role of $C_{60}$ in Xe@$C_{60}$ photoionization is twofold. $C_{60}$ acts as a potential resonator that reflects the photoelectron wave coming from Xe atom. This leads to oscillations in the frequency dependence of the photoionization cross sections [4].

The $C_{60}$ shell at some frequencies acts as a dynamical screen that is capable to suppress or enhance the incident electromagnetic radiation acting upon the doped atom A [7-9]. This effect is due to dynamical polarization of the collectivized electrons of the fullerene shell. Plasma excitations of these electrons generate the alternating dipole moment. This dipole moment causes the ionization of the electronic shells of the endohedral atom. The screening effects of the $C_{60}$ shell are particularly strong for incident radiation frequency $\omega$ of about $C_{60}$ Giant resonance, i.e. 20 – 22 eV[2].

We will show in this Letter that the dynamic polarization of $C_{60}$ increases the outer shell photoionization cross-section at any $\omega$. The maximal enhancement is in the region of $C_{60}$ dipole polarizability maximum. At the same time, the angular anisotropy is affected only by $C_{60}$ reflection and not by its polarization.

---

[1] As a first example of such a research, let us mention the tentative data on measurements of photoionization cross-section of Ce@$C_{82}$ [6].
[2] Atomic system of units is used in this paper

The resonator and dynamic screen effects of $C_{60}$ manifest themselves together by transforming a smoothly decreasing $5p^6$ cross-section in Xe into a prominent maximum that we call Giant Endohedral Resonances (GER) for Xe@$C_{60}$. Its oscillator strength is as big as 25! We believe that similar is the situation for other endohedrals Xe@F with F = $C_{70}$, $C_{76}$, $C_{82}$, $C_{87}$ etc

**2.** We employ the theoretical approaches already developed in a number of previous papers (see [8, 9] and references therein). However, for completeness, we will repeat here the main points of the consideration and present the main formula used in calculations.

We will start with the problem of an isolated closed shell atom. The following relation gives the differential in angle photoionization cross-section by non-polarized light in the dipole approximation [10]:

$$\frac{d\sigma_{nl}(\omega)}{d\Omega} = \frac{\sigma_{nl}(\omega)}{4\pi}[1 - \frac{\beta_{nl}}{2}P_2(\cos\theta)], \tag{1}$$

where $\kappa = \omega/c$, $P_l(\cos\theta)$ are the Legendre polynomials, $\theta$ is the angle between photon $\kappa$ and photoelectron velocity **v**, $\beta_{nl}(\omega)$ is the dipole angular anisotropy parameter.

There are two possible dipole transitions from subshell $5p^6$, namely $p \rightarrow d, s$, the amplitudes of which in one-electron Hartree-Fock (HF) approximation are denoted as $d_d$ and $d_s$, respectively. The following relation gives the cross-section $\sigma_{np}(\omega)$:

$$\sigma_{np}(\omega) = 8\pi^2 \omega c^{-1}[2d_d^2 + d_s^2], \tag{2}$$

while the parameter $\beta_{np}(\omega)$ looks as

$$\beta_{np}(\omega) = 2[d_d^2 - 2d_d d_s \cos(\delta_d - \delta_s)](2d_d^2 + d_s^2)^{-1}. \tag{3}$$

Here $\delta_l(k)$ are the photoelectrons' scattering phases; the following relation gives the matrix elements $d_{d,s}$ in the so-called $r$-form

$$d_{d,s} \equiv \int_0^\infty P_{np}(r) r P_{\varepsilon d,s}(r) dr, \tag{4}$$

where $P_{np}(r)$, $P_{\varepsilon f,d}(r)$ are the radial HF one-electron wave functions of the $np$ discrete level and $\varepsilon d, s$ - in continuous spectrum, respectively [10].

In order to take into account the Random Phase Approximation with Exchange (RPAE) [10] multi-electron correlations, one has to perform the following substitutions in the expressions for $\beta_{np}(\omega)$:

$$d_d d_s \cos(\delta_d - \delta_s) \rightarrow \tilde{D}_d \tilde{D}_s \cos(\delta_d + \Delta_d - \delta_s - \Delta_s); \quad D_{d,s}(\omega) \equiv \tilde{D}_{d,s}(\omega)\exp[i\Delta_{d,s}(\varepsilon)]. \tag{5}$$

The following RPAE equation [10] for dipole amplitude has to be solved numerically [11]:

$$\langle v_2|D(\omega)|v_1\rangle = \langle v_2|d|v_1\rangle + \sum_{v_3,v_4} \frac{\langle v_3|D(\omega)|v_4\rangle(n_{v_4}-n_{v_3})\langle v_4 v_2|U|v_3 v_1\rangle}{\varepsilon_{v_4}-\varepsilon_{v_3}+\omega+i\eta(1-2n_{v_3})}, \qquad (6)$$

where

$$\langle v_1 v_2|\hat{U}|v'_1 v'_2\rangle \equiv \langle v_1 v_2|\hat{V}|v'_1 v'_2\rangle - \langle v_1 v_2|\hat{V}|v'_2 v'_1\rangle. \qquad (7)$$

Here $\hat{V}\equiv 1/|\vec{r}-\vec{r}'|$ and $v_i$ is the total set of quantum numbers that characterize a HF one-electron state on discrete (continuum) levels. That includes the principal quantum number (energy), angular momentum, its projection and the projection of the electron spin. The function $n_{v_i}$ (the so-called step-function) is equal to 1 for occupied and 0 for vacant states.

**3.** The confinement effects that take place near the photoionization threshold can be described within the framework of the "orange" skin potential model (see e.g. [12]). According to this model, for small photoelectron energies the real static potential of the $C_{60}$ can be presented as a zero-thickness, bubble-type pseudo-potential [12] with radius $R$ and strength $V_0$. The parameter $V_0$ is determined by the requirement that the binding energy of the extra electron in the negative ion $C_{60}^-$ is equal to its observable value. Addition of the fullerene potential to the atomic potential leads to a factor $F_l(k)$ in the photoionization amplitudes which depends only upon the photoelectron's momentum $k$ and orbital quantum number $l$ [11]:

$$F_l(k) = \cos\Delta_l(k)[1-\tan\Delta_l(k)v_{kl}(R)/u_{kl}(R)], \qquad (8)$$

where $\Delta_l(k)$ are the additional phase shifts due to the fullerene shell potential. They are expressed by the following formula:

$$\tan\Delta_l(k) = u_{kl}^2(R)/[u_{kl}(R)v_{kl}(R)+k/2V_0]^{-1}. \qquad (9)$$

In (8, 9) $u_{kl}(r)$ and $v_{kl}(r)$ are the regular and irregular solutions of the atomic HF equations for a photoelectron with momentum $k=\sqrt{2\varepsilon}$, where $\varepsilon$ is the photoelectron energy connected with the photon energy $\omega$ by the relation $\varepsilon = \omega - I_A$ with $I$ being the atom A ionization potential.

The following relations connect the amplitudes $D^{AC(r)}$ and cross-sections for endohedral atom A@$C_{60}$ with the respective values for isolated atoms that correspond to $nl\to\varepsilon l'$ transitions:

$$D^{AC(r)}_{nl,kl'}(\omega) = F_{l'}(k)D_{nl,kl'}(\omega); \quad \sigma^{AC}_{nl,kl'}(\omega) = F_l^2(k)\sigma^A_{nl,kl'}(\omega). \qquad (10)$$

Note that index ($r$) marks that photoelectron's reflection is taken into account.

The effect of the fullerene electron shell polarization upon atomic photoionization amplitude can be taken into account in RPAE. An essential simplification comes from the facts that the $C_{60}$ radius $R_c$ is big enough, $R_c \gg r_s$, and the electrons in $C_{60}$ are located within a layer, the thickness of which $\Delta R_c$ is considerably smaller than $R_c$. In this case the amplitude of endohedral atom's photoionization with all essential atomic correlations taken into account can be presented by the following formula:

$$D_{nl,nl'}^{AC(r)}(\omega) \cong F_{l'}(k) D_{nl,nl'}^{A}(\omega)\left(1 - \alpha_{C_{60}}(\omega)/R_{C_{60}}^{3}\right) \equiv F_{l'}(k) G_{C_{60}}(\omega) D_{nl,nl'}^{A}(\omega), \qquad (11)$$

where $\alpha_{C_{60}}(\omega)$ is the dipole dynamical polarizability of $C_{60}$ and $R_{C_{60}}$ is its radius.

Using the relation between the imaginary part of the polarizability and the dipole photoabsorption cross-section $\mathrm{Im}\,\alpha_{C_{60}}(\omega) = c\sigma_{C_{60}}(\omega)/4\pi\omega$ one can derive the polarizability of the $C_{60}$ shell. Although experiments [13, 14] do not provide absolute values of $\sigma_{C_{60}}^{d}(\omega)$, it can be reliably estimated using different normalization procedures on the basis of the sum rule: $(c/2\pi^{2})\int_{I_{o}}^{\infty}\sigma_{C_{60}}(\omega)d\omega = N$, where $N$ is the number of collectivized electrons. The real part of polarizability is connected with imaginary one (and with the photoabsorption cross-section) by the dispersion relation:

$$\mathrm{Re}\,\sigma_{C_{60}}(\omega) = \frac{c}{2\pi^{2}} \int_{I_{60}}^{\infty} \frac{\sigma_{C_{60}}(\omega')d\omega'}{\omega'^{2} - \omega^{2}}, \qquad (12)$$

where $I_{60}$ is the $C_{60}$ ionization potential. This approach was used for polarizability of $C_{60}$ in [9], where it was considered that $N = 240$, i.e. 4 collectivized electrons per each C atom in $C_{60}$. Using the photoabsorption data that are considered as most reliable in [13], we obtained $N_{eff} \approx 250$ that is sufficiently close to the value, assumed in [9].

Using the amplitude (11), one has for the cross section

$$\sigma_{5p \to kd,s}^{AC(r)}(\omega) = \sigma_{5p \to kd,s}^{A}(\omega) F_{d,s}^{2}(k)\left|1 - \alpha_{C_{60}}^{d}(\omega)/R_{C_{60}}^{3}\right|^{2} \equiv F_{d,s}^{2}(k) S(\omega) \sigma_{5d \to kd,s}^{A}(\omega), \qquad (13)$$

where the fullerenes polarization factor $S(\omega)$ is defined as

$$S(\omega) = \left|1 - \alpha_{C_{60}}^{d}(\omega)/R_{C_{60}}^{3}\right|^{2}. \qquad (14)$$

**4.** The $C_{60}$ parameters in the present calculations were chosen the same as in the previous papers, e.g. in [8]: $R = 6.639$ and $V_{0} = 0.443$. In Fig. 1 we present the factor $S(\omega)$ for fullerene $C_{60}$ [8], where arrows mark the $5p^{6}$ ionization thresholds. The sophisticated behavior of $S(\omega)$ as a function of $\omega$ along with the effect from photoelectron's reflection leads to formation of prominent resonances.

In Fig. 2 we present the photoionization cross-section of 5p electrons of Xe@$C_{60}$. We have observed three maxima, connected to the reflection $F_{l}(k)$ factors, but strongly enhanced by the factor $S(\omega)$.

In Fig. 3 we present the $\beta_{np}(\omega)$ parameter for Xe and Xe@$C_{60}$ As is seen, this parameter is strongly affected by photoelectron reflection but is untouched by the $C_{60}$ polarization that is the same for both, $p \to d$ and $p \to s$, transitions. It is remarkable that the oscillatory structure in $\beta_{np}(\omega)$ is more pronounced than in the cross-section, since $\beta_{np}(\omega)$ is determined by the *ratios* of two differently oscillating matrix elements: $F_{d}(k)D_{np,kd}(\omega)$ and $F_{s}(k)D_{np,ks}(\omega)$.

The photoionization cross-section of 5p electrons in Xe@$C_{60}$ impressively exceeds the photoionization cross-section of isolated Xe atom. The monotonically decreasing curve in the isolated atom is transformed into a curve with a big maximum. It is remarkable that the total oscillator strength of the big maximum is about 25, i.e. 2.5 times bigger than that of the 4d atomic Giant resonance in isolated Xe. As is known, the 4d-resonance is the strongest in isolated atoms. Due to this increase of the "caged" atom photoionization cross-section, the total oscillator strength sum in the $\omega$ range, say, $I < \omega < I + 1Ry$ increases dramatically.

A natural question is: what is the origin of this increase? The answer is as follows – this increase comes from the fullerene shell, and not due to redistribution of the "caged" atom oscillator strength. The latter is evident, since this atom, simply speaking, has not enough electrons for this. Note that the total sum rule for an endohedral is equal to $N_F + N_A$, where $N_F$ and $N_A$ are the total numbers of electrons in the fullerene and the atom, respectively. As to the sum of oscillator strengths of the "caged" Xe atom, it increases, roughly speaking, by the area on the Fig. 2 that is between the solid and dashed-dotted curves.

The maximum in Fig. 2 is reasonable to call Giant Endohedral resonance. No doubt that similar effect takes place in outer shells of endohedral atoms and for atoms "caged" by other than $C_{60}$ fullerenes.

There is no doubt that such resonances can be detected in experimental studies of photoionization of endohedral atoms, using the photoelectron spectroscopy method.

**Acknowledgement**

MYaA acknowledges financial support by the Israeli Science Foundation, Grant 174/03 and the Hebrew University Intramural Funds. ASB expresses his gratitude to the Hebrew University for hospitality and for financial support by Uzbekistan National Foundation, Grant ФА-Ф4- Ф095.

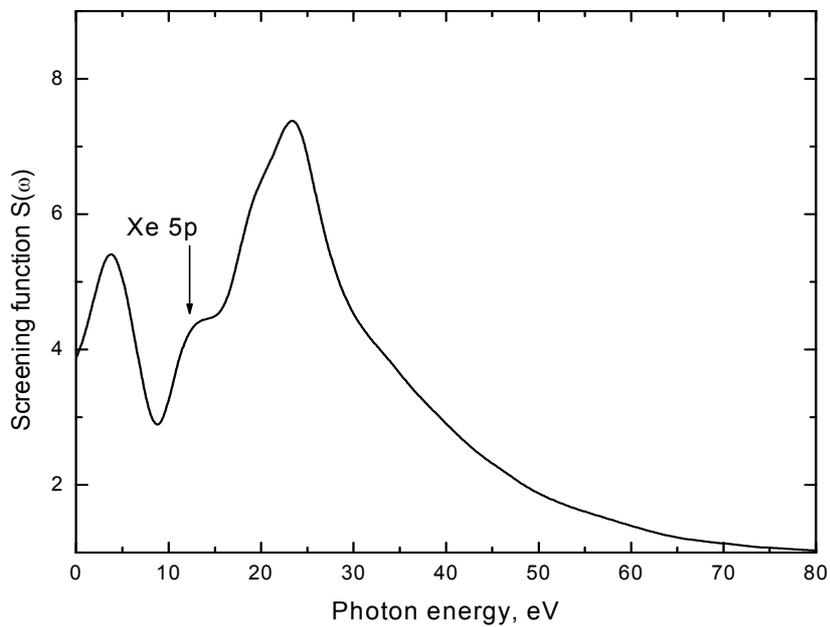

Fig.1. Enhancement factor $S(\omega)$ due to $C_{60}$ dipole polarization by light beam

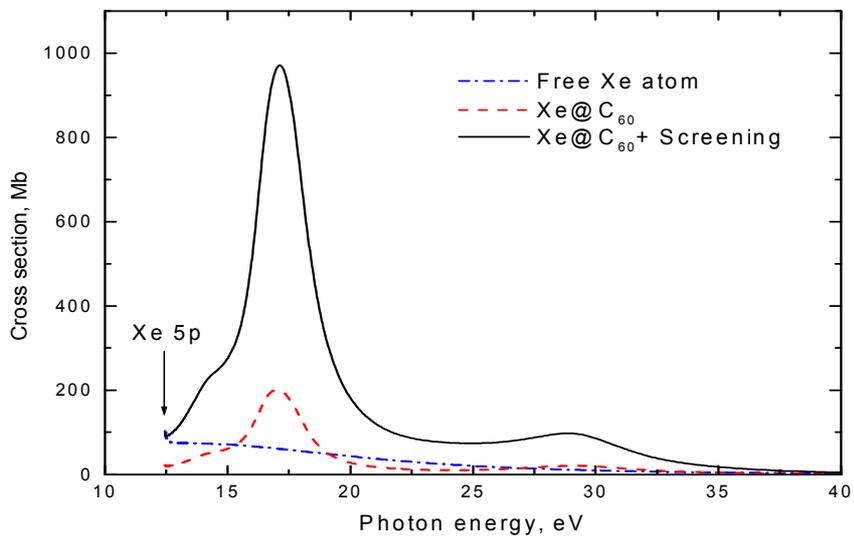

Fig.2. Photoionization cross-sections of $5p$ electrons in Xe@$C_{60}$
- solid line; dashed line –with account of $C_{60}$ reflection; dashed-dotted line – isolated atom

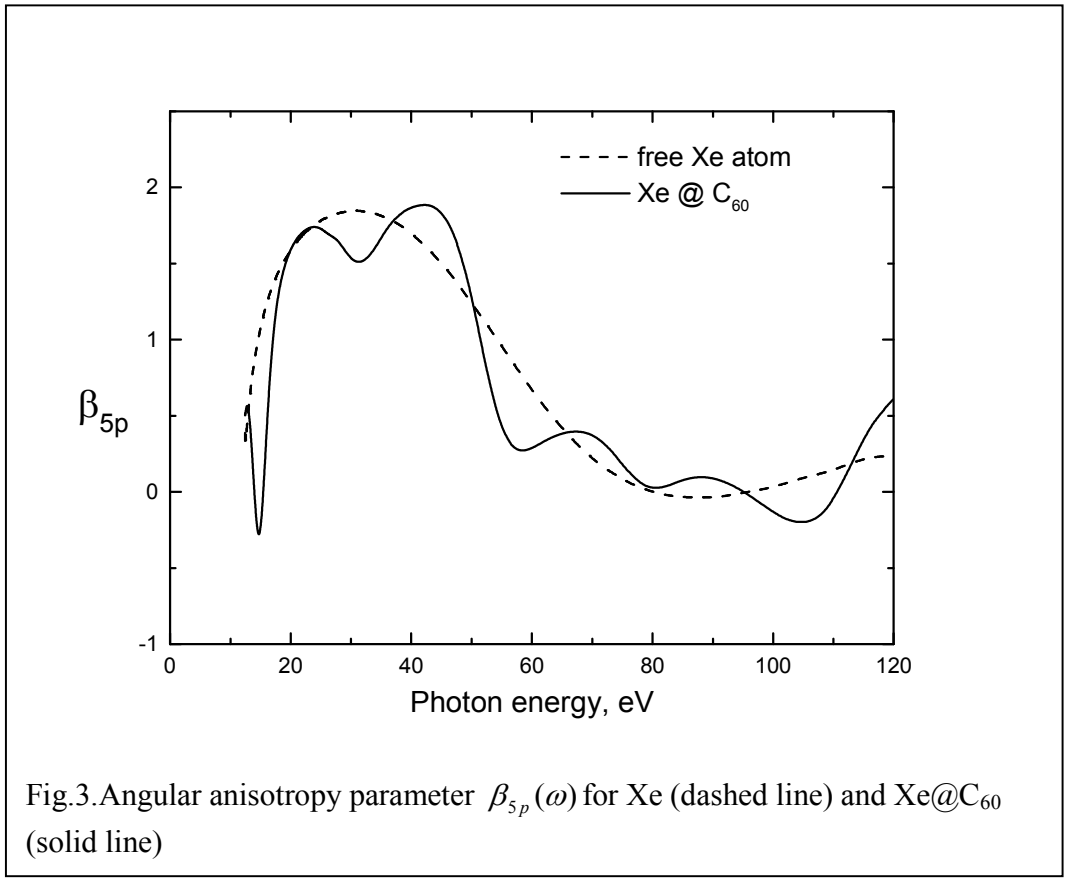

Fig.3. Angular anisotropy parameter $\beta_{5p}(\omega)$ for Xe (dashed line) and Xe@$C_{60}$ (solid line)